\documentclass[letterpaper,aps,prd,onecolumn,showpacs,preprintnumbers,nofootinbib]{revtex4}
\usepackage{graphicx}
\usepackage[intlimits,centertags]{amsmath}
\usepackage{amssymb,amsfonts}
\usepackage{gensymb}
\usepackage{mathrsfs}
\usepackage{hyperref}

\begin{document}

\title{Gamma-ray halos as a measure of intergalactic magnetic fields:\\ a classical moment problem}

\author{Markus~Ahlers} 
\affiliation{C.N.Yang Institute for Theoretical Physics, SUNY at Stony Brook, Stony Brook, NY 11794-3840, USA}

\begin{abstract}
The presence of weak intergalactic magnetic fields can be studied by their effect on electro-magnetic cascades induced by multi-TeV $\gamma$-rays in the cosmic radiation background. Small deflections of secondary electrons and positrons as the cascade develops extend the apparent size of the emission region of distant TeV $\gamma$-ray sources. These $\gamma$-ray halos can be resolvable in imaging atmospheric Cherenkov telescopes and serve as a measure of the intergalactic magnetic field strength and coherence length. We present a method of calculating the $\gamma$-ray halo for isotropically emitting sources by treating magnetic deflections in the cascade as a diffusion process. With this ansatz the moments of the halo follow from a set of simple diffusion-cascade equations. The reconstruction of the angular distribution is then equivalent to a classical moment problem. We present a simple solution using Pad\'e approximations of the moment's generating function.
\end{abstract}


\preprint{YITP-SB-11-10}

\maketitle

\section{Introduction}\label{sec:introduction}

The presence of large-scale magnetic fields in cosmic environments can be probed by various astronomical techniques. Synchrotron radiation of relativistic electrons can be detected by its characteristic linear polarization and spectrum. Faraday rotation of linearly polarized emission tests the birefringent properties of a dilute magnetized plasma filling intergalactic space. Zeeman splitting of an atom's energy levels can be observed by the corresponding shift of spectral lines from astrophysical masers. With these standard methods it has been possible to identify micro-Gauss magnetic fields coherent over galactic scales in many galaxies and galaxy clusters~\cite{Kronberg:1993vk,Beck:2008ty}.

The origin of these large-scale magnetic fields is unclear. It is assumed that galactic magnetic fields can be maintained and amplified via a dynamo mechanism, where the kinetic energy of a turbulent interstellar plasma is converted into magnetic energy~\cite{Kulsrud:2007an}. However, this requires initial seed fields of unknown origin, possibly pre-galactic or primordial~\cite{Grasso:2000wj,Widrow:2002ud}. The strength and correlation length of primordial intergalactic magnetic fields (IGMFs) can be limited by their effect on various stages in cosmic history. The strongest bounds on the strength of primordial IGMFs arise from the study of temperature anisotropies in the cosmic microwave background (CMB)~\cite{Barrow:1997mj}. The limits are of the order of nano-Gauss for a correlation lengths larger than a few Mpc. Simulations of large-scale structure formation favor long-range IGMF with a strength of the order of pico-Gauss~\cite{Dolag:2004kp}.

It has been suggested that weak IGMFs of the order of fempto-Gauss can be probed by their effect on electro-magnetic cascades initiated by the emission of distant multi-TeV $\gamma$-ray sources~\cite{Aharonian:1993vz,Plaga:1995}. High-energy $\gamma$-rays produce pairs of electrons/positrons in the cosmic infrared/optical background (CIB) with an interaction length of the order of 100~Mpc. The secondary leptons lose their energy via inverse-Compton scattering off the background photons and produce secondary $\gamma$-rays at somewhat lower energies. If these photons are still above the pair-production threshold the cycle repeats. In this way the electro-magnetic energy of the cascade is continuously shifted into the GeV-TeV energy region. In the presence of magnetic fields secondary leptons are deflected off the line-of-sight and secondary $\gamma$-rays inherit this deflection. This will attenuate the flux originally emitted towards the observer. However, $\gamma$-rays initially emitted away from the observer can be scattered back into the line-of-sight and partially compensate for this loss.

There are various ways to infer the strength $B_0$ and correlation length $\lambda_B$ of the IGMF from this effect. For small deflections and isotropically emitting sources (or sufficiently large jet opening-angles) the net effect will be an extended emission region of secondary $\gamma$-rays~\cite{Aharonian:1993vz,Neronov:2007zz,Eungwanichayapant:2009bi,Elyiv:2009bx,Dolag:2009iv}. For burst-like $\gamma$-ray sources this can also cause an observable time-delay between the primary burst and secondary $\gamma$-rays~\cite{Plaga:1995,Murase:2008pe}. In the case of a hard TeV $\gamma$-ray emission the secondary component can dominate over the attenuated primary $\gamma$-rays. Non-observation of the point-source in the GeV-TeV band can then imply a lower limit on the magnetic field depending on the instrument's resolution~\cite{Aharonian:1993vz}. These methods have been applied to various TeV $\gamma$-ray sources~\cite{Neronov:1900zz,d'Avezac:2007sg,Tavecchio:2010mk,Dolag:2010ni,Dermer:2010mm,Taylor:2011bn,Neronov:2011ni} and indicate the presence of an IGMF. The inferred lower limits on its strength range from $10^{-18}$~G to $10^{-15}$~G, depending on many systematic uncertainties like the primary emission spectrum, the CIB and the coherence length of magnetic fields.

Besides the systematic uncertainties of these methods, there are also some technical challenges in calculating the energy and angular spectrum of the $\gamma$-ray halos. A straightforward Monte-Carlo calculation of the electro-magnetic cascade can become numerically expensive; since energy is conserved in the cascade the number of $\gamma$-rays, electrons and positrons in the cascade increases by one order of magnitude for every decade of the energy shift. At every step of the cascade each particle will have accumulated a deflection angle with respect to the line-of-sight which depends on its history in the cascade. In order to accumulate a satisfactory resolution in energy and angular extend of the halos it is necessary to sample over many cascades.

In the absence of deflections by magnetic fields the electro-magnetic cascade can be calculated efficiently by analytical methods using cascade equations and the method of matrix doubling~\cite{Protheroe:1992dx}. We will show in this paper that there is a straightforward extension of this method to the case of isotropic emitters and small deflections in magnetic fields. The key observation is that the deflection $\theta$ of electrons and positrons in the cascade in combination with inelastic losses to photons can be treated as a diffusion process in $\theta$-space where the diffusion coefficient depend on the particle's Larmor radius and the inverse-Compton energy loss length. We derive diffusion-cascade equations that describe the evolution of the moments of the $\theta$-distribution and give a simple method how these moments can be used to reconstruct the distribution.

We will begin in section~\ref{sec:cascades} by a discussion of electro-magnetic cascades from $\gamma$-ray point sources in the presence of weak IGMFs. In section~\ref{sec:diffusion} we will motivate the extension of the Boltzmann equations by a diffusion term in $\theta$-space and give an extended set of cascade equations for the moments of the $\theta$-distribution. We discuss in section~\ref{sec:reconstruction} how the full $\theta$-distribution can be reconstructed efficiently from a finite number of moments via explicit inverse Laplace transformations of Pad\'e approximations of the moment's generating function. We will test our method in section~\ref{sec:example} by two examples and compare our results to previous studies. Finally, we conclude in section~\ref{sec:conclusion}. 

We work throughout in natural Heaviside-Lorentz units with $\hbar=c=\epsilon_0=\mu_0=1$, $\alpha=e^2/(4\pi)\simeq1/137$ and $1~{\rm G} \simeq 1.95\times10^{-2}{\rm eV}^2$. 

\section{Electro-magnetic Cascades}\label{sec:cascades}

The driving processes of the electro-magnetic cascade in the cosmic radiation background (CRB) are inverse Compton scattering (ICS) with CMB photons, $e^\pm+\gamma_{\rm bgr}\to e^\pm+\gamma$, and pair production (PP) with CMB and CIB radiation, $\gamma+\gamma_{\rm bgr}\to e^++e^-$~\cite{Blumenthal:1970nn,Blumenthal:1970gc}. In particular, the interaction length of multi-TeV $\gamma$-rays depend on the CIB background at low redshift and is of the order of a few 100~Mpc. We show the relevant interaction lengths and energy loss lengths in the left panel of Fig.~\ref{fig:fig1}. High energetic electrons and positrons may also lose energy via synchrotron radiation in the intergalactic magnetic field, but this contribution is in general negligible for the small magnetic field strength considered here. Further processes contributing to the electro-magnetic cascade are double pair production, $\gamma+\gamma_{\rm bgr}\to e^++e^-+e^++e^-$, and triple pair production, $e^\pm+\gamma_{\rm bgr}\to e^\pm+e^++e^-$~\cite{Protheroe:1992dx,Lee:1996fp}. These contributions can be neglected for cascades initiated by multi-TeV $\gamma$-rays considered here. Also, interactions on the cosmic radio background are negligible in this case.

For the calculation of the flux from a $\gamma$-ray point-source it is convenient to start from the evolution of a comoving number density $Y_\alpha = n_\alpha/(1+z)^3$ (GeV${}^{-1}$ cm${}^{-1}$) in a spatially homogeneous and isotropic universe. The Boltzmann equations of electrons/positrons ($Y_e$) and $\gamma$-rays ($Y_\gamma$) is given by
\begin{equation}\label{eq:boltzmann}
\dot {Y}_\alpha(E) = \partial_E(HE{Y}_\alpha)- \Gamma_\alpha{Y}_{\alpha}(E)+\sum_{\beta=e,\gamma}\int_E{\rm d}E'\gamma_{\beta\alpha}(E',E){Y}_{\beta}(E')+\mathcal{L}_\alpha(E)\,,
\end{equation}
together with the Friedman-Lema\^{\i}tre equations describing the cosmic expansion rate $H(z)$ as a function of the redshift $z$, $H^2 (z) = H^2_0\,[\Omega_{\rm m}(1 + z)^3 + \Omega_{\Lambda}]$, normalized to its present value of $H_0 \sim70$ km\,s$^{-1}$\,Mpc$^{-1}$. We consider the usual ``concordance model'' dominated by a cosmological constant with $\Omega_{\Lambda} \sim 0.7$ and a (cold) matter component, $\Omega_{\rm m} \sim 0.3$~\cite{Nakamura:2010zzi}.  The time-dependence of the redshift is given by ${\rm d}z = -{\rm d} t\,(1+z)H$. The first term in the r.h.s.~of Eq.~(\ref{eq:boltzmann}) accounts for the continuous energy loss due to the adiabatic expansion of the Universe. The second and third terms describe the interactions with background photon fields involving particle losses ($\alpha \to$ anything)  and particle generation $\alpha\to\beta$. The angular-averaged (differential) interaction rate, $\Gamma_\alpha$
($\gamma_{\alpha\beta}$) is defined as
\begin{gather}\label{eq:diffgamma1}
\Gamma_{\alpha}(z,E_\alpha) =
\frac{1}{2}\int_{-1}^1{\rm d}\cos\theta\int{\rm d}
\epsilon\,(1-\beta
\cos\theta) n_\gamma(z,\epsilon)\sigma^{\rm tot}_{\alpha\gamma}\,,\\
\gamma_{\alpha\beta}(z,E_\alpha,E_\beta) = \Gamma_\alpha(z,E_\alpha)\,\frac{{\rm d} N_{\alpha\beta}}{{\rm d} E_\beta}(E_\alpha,E_\beta)\,,\label{eq:diffgamma2}
\end{gather}
where $n_\gamma(z,\epsilon)$ is the energy distribution of background photons at redshift $z$ and ${\rm d} N_{\alpha\beta}/{\rm d} E_\beta$ is the angular-averaged distribution of particles $\beta$ after interaction of a particle $\alpha$. Besides the contribution of the CMB we use the CIB from Ref.~\cite{Franceschini:2008tp}. Due to the cosmic evolution of the radiation background density the interaction rates (\ref{eq:diffgamma1}) and (\ref{eq:diffgamma2}) scale with redshift. The CMB evolution follows an adiabatic expansion, $n_\gamma(z,\epsilon) = (1+z)^2\,n_\gamma(0,\epsilon/(1+z))$, and we assume the same evolution of the CIB for simplicity. We refer to Ref.~\cite{Ahlers:2009rf} for a list of the redshift scaling relations of the interaction rates in Eqs.~(\ref{eq:boltzmann}). The last term in Eq.~(\ref{eq:boltzmann}), $\mathcal{L}_\alpha$, accounts for the emission rate of particles $\alpha$ per comoving volume. 

In the limit of small deflections of particles via magnetic fields, the flux from a $\gamma$-ray point source at redshift distance $z^\star$ with emission rate $Q_\gamma$ is equivalent to an angular-averaged flux from a sphere at redshift $z^\star$. Hence, the solution of $Y$ at $t=0$ is equivalent to the point source flux $J$ (GeV${}^{-1}$ cm${}^{-2}$ s${}^{-1}$) by replacing the emission rate density $\mathcal{L}$ in (\ref{eq:boltzmann}) by
\begin{equation}\label{eq:PS}
\mathcal{L}_\gamma^\star(z,E) = \frac{Q_\gamma(E)}{4\pi d_C^2(z^\star)}H(z^\star)\delta(z-z^\star)\,,
\end{equation}
where the comoving distance of the source (in a flat universe) is given by $d_C(z) \equiv \int_0^z{\rm d}z'/H(z')$. Note, that we can also use the ansatz (\ref{eq:PS}) for a cosmic ray (CR) point source located at redshift $z^\star$, where the electro-magnetic emission is in the form of cosmogenic $\gamma$-rays, electrons and positrons produced during CR propagation~\cite{AhlersSalvado}.

\section{Angular Diffusion in Intergalactic Magnetic Fields}\label{sec:diffusion}

The $\gamma$-ray cascade can only contribute to a GeV-TeV point-source flux if the deflections of secondaries off the line-of-sight are sufficiently low. The scattering angle of secondaries is only of the order of $\epsilon/m_e$ and can be neglected for the optical/infra-red background photon energies $\epsilon$. However, electrons and positrons can be deflected in the IGMF. We can estimate the extend of the cascaded $\gamma$-ray emission by simple geometric arguments following~\cite{Neronov:2007zz}. Deflection of electrons and positrons will be small if the energy loss length $\lambda_e$ of electrons/positrons via inverse Compton scattering (ICS) is much smaller than the Larmor radius given as $R_L = E/eB \simeq{1.1} (E_{\rm TeV}/B_{\rm fG}){\rm Mpc}$. Here and in the following we use the abbreviations $E = E_{\rm TeV}{\rm TeV}$, etc. For center of mass energies much lower than the electron mass, corresponding to energies below PeV in the CMB frame, electrons and positrons interact quickly on kpc scales but with low inelasticity proportional to their energy, $\lambda_e\simeq 0.4~{\rm Mpc}/E_{\rm TeV}$. The typical size of the point-spread function (PSF) of imaging atmospheric Cherenkov telescopes (IACTs) is of the order of $\theta_{\rm PSF}\simeq0.1^\circ$. Hence, magnetic deflections become important if $\theta_{\rm PSF}\lesssim \lambda_e/R_L$ or $E\lesssim 14~{\rm TeV} \sqrt{B_{\rm fG}/\theta_{\rm PSF, 0.1^\circ}}$. 

In the following we are going to study these magnetic deflections more quantitatively. For simplicity, we will start with a regular IGMF that fills the space between the source and the observer and has the component $B_\perp$ perpendicular to the line-of-sight. We also assume that the source is emitting $\gamma$-rays isotropically.\footnote{We can relax this condition by requiring that the $\gamma$-ray emission is into a jet with sufficiently large jet opening-angle.} Due to charge conservation in the cascade electrons and positrons will be produced in equal rates and will be deflected in opposite directions. For small scattering and isotropic emission we can assume that leptons that are lost by deflections out of the line-of-sight are replenished by the corresponding leptons deflected into the line-of-sight. Effectively, we can hence assume that the total number of electrons/positrons within the line-of-sight remains constant by these deflections while the scattering angle $\theta$ is broadened by the magnetic field. The width of this $\theta$-distribution, ${\mathcal Y}_e(E,\theta)$, is determined by the energy loss length via ICS. Secondary $\gamma$-rays will inherit the $\theta$-distribution of the parent leptons and will appear as extended halos.

The energy loss length via ICS with CMB photons is much smaller than the typical distance of TeV $\gamma$-ray sources or the interaction length of PP in the CIB. This indicates that we can treat magnetic deflections in the cascade as a diffusive process of the angle $\theta$. The mean free path of the electrons/positrons corresponds to the energy loss rate in ICS and the diffusion velocity is the inverse Larmor radius. Hence, the diffusion coefficient $D$ is of the order of $\lambda^2_{\rm ICS}/R^2_L$. A more rigorous derivation (see Appendix~\ref{sec:appI}) shows that the evolution of the $\theta$-distributions of leptons and $\gamma$-rays, ${\mathcal Y}_e(E,\theta)$ and ${\mathcal Y}_\gamma(E,\theta)$ respectively, can be described by the coupled set of differential equations,
\begin{align}\label{eq:diffregular}
\dot  {\mathcal Y}_\gamma&\simeq  \partial_E(HE{\mathcal Y}_\gamma)-\Gamma_\gamma {\mathcal Y}_\gamma + \sum_{\alpha=e,\gamma}\int_E{\rm d}E' \gamma_{\alpha \gamma}(E',E) {\mathcal Y}_\alpha(E') + \mathcal{L}^\star_\gamma\delta(\theta)\,,\\
\dot  {\mathcal Y}_e &\simeq \partial_E(HE{\mathcal Y}_e)-\Gamma_e {\mathcal Y}_e  + \sum_{\alpha=e,\gamma}\int_E{\rm d}E' \gamma_{\alpha e}(E',E) {\mathcal Y}_\alpha(E')  + \mathcal{L}^\star_e\delta(\theta)+ \int\limits_E{\rm d}E'{\mathcal D}_{\rm reg}(E',E)\partial^2_{\theta} {\mathcal Y}_e(E')\,.
\end{align}
The diffusion matrix of electrons/positrons in a regular magnetic field is given by 
\begin{equation}\label{eq:Dregular}
{\mathcal D}_{\rm reg}(E',E) = \frac{1}{E\,\Gamma_{\rm ICS}(E)}\frac{e^2B_\perp^2}{E'^2\,\langle x\rangle(E')}\,,
\end{equation}
where $\langle x\rangle(E)$ is the inelasticity of ICS with interaction rate $\Gamma_{\rm ICS}(E)$. For cosmological sources the redshift scaling of the diffusion matrix (\ref{eq:Dregular}) can also become important. For primordial magnetic fields scaling as $B_\perp(z) = (1+z)^2B_\perp(0)$ and ICS with CMB photons the redshift dependence is given by the simple relation ${\mathcal D}_{\rm reg}(z,E',E) = (1+z)^4{\mathcal D}_{\rm reg}(0,(1+z)E',(1+z)E)$.

This formalism has the advantage that we can calculate the moments of the $\theta$-distribution by an extended set of cascade equations as we will see in the following. Firstly, we introduce the quantities
\begin{equation}\label{eq:Thetan}
{Y}_{e/\gamma}^{(n)} \equiv \frac{1}{(2n)!}\int\limits_{-\infty}^\infty{\rm d}\theta\,\theta^{2n}\,{\mathcal Y}_{e/\gamma}\quad \text{(regular)}\,.
\end{equation}
At leading order we have ${Y}^{(0)} = Y$ as the solution of Eq.~(\ref{eq:boltzmann}) and for $n\geq1$ the quantities ${Y}^{(n)}$ correspond to the scaled moments of the $\theta$-distribution.\footnote{To be more precise, $\theta$ is an element of the covering space $\mathbf{R}$ of the circle $\mathbf{S}^1$. The distribution along the circle is then obtained by ${\mathcal Y}_{\mathrm{S}^1}(E,\theta) = \sum_{n\in\mathrm{Z}}{\mathcal Y}_{\mathrm{R}}(E,\theta+2\pi n)$. However, we are only interested in small scattering angles $\theta\ll 1^\circ$ and hence ${\mathcal Y}_{\mathrm{S}^1}(E,\theta) \simeq {\mathcal Y}_{\mathrm{R}}(E,\theta)$.} It is easy to see that the quantities ${Y}^{(n)}$ ($n>0$) follow the coupled set of differential equations,
\begin{equation}\label{eq:Thetanevol}
\dot {Y}^{(n)}_\alpha(E) = \partial_E(HE{Y}^{(n)}_\alpha)- \Gamma_\alpha{Y}_{\alpha}^{(n)}(E)+\sum_{\beta=e,\gamma}\int_E{\rm d}E'\gamma_{\beta\alpha}(E',E){Y}_{\beta}^{(n)}(E')
+ \delta_{e\alpha}\int_E{\rm d}E'{\mathcal D}(E',E){Y}_\alpha^{(n-1)}(E')\,,
\end{equation}
in addition to Eqs.~(\ref{eq:boltzmann}). Note, that electro-magnetic interactions of photons and leptons that drive the cascade happen on time-scales much shorter than the rate of adiabatic losses in the Universe. We can hence treat the interaction rates as constant over small time-intervals and neglect the energy loss terms $\partial_E(HE{Y}_\alpha^{(n)})$ in Eqs.~(\ref{eq:boltzmann}) and (\ref{eq:Thetanevol}). We show in Appendix~\ref{sec:appII} that this system of equations can then be solved efficiently by a generalization of the conventional cascade equations. 

We next consider a randomly oriented IGMF field with a coherence length $\lambda_B$ much smaller then the distance $d$ of the source. In this case we have to replace Eq.~(\ref{eq:diffregular}) by the evolution of radial diffusion on a sphere of the form\footnote{We consider only small deflections and can hence approximate the sphere as two-dimensional flat space.}
\begin{equation}\label{eq:diffrandom}
\dot  {\mathcal Y}_e\simeq \partial_E(HE{\mathcal Y}_e)-\Gamma_e {\mathcal Y}_e + \sum_{\alpha=e,\gamma}\int_E{\rm d}E' \gamma_{\alpha e}(E',E) {\mathcal Y}_\alpha(E')  + \mathcal{L}^\star_\gamma\delta(\theta) + \int\limits_E{\rm d}E'{\mathcal D}_{\rm rnd}(E',E)\theta^{-1}\partial_{\theta}\left[\theta\partial_\theta {\mathcal Y}_e(E')\right] \,,
\end{equation}
with diffusion coefficient (see Appendix~\ref{sec:appI})
\begin{equation}\label{eq:Drandom}
{\mathcal D}_{\rm rnd}(E',E) \simeq \frac{1}{3}\frac{\min(1,\lambda_B\Gamma_{\rm ICS}(E))}{E\,\Gamma_{\rm ICS}(E)}\frac{e^2B_0^2}{E'^2\langle x\rangle(E')}\,.
\end{equation}
Here, a factor $1/3$ accounts for the random orientation of the magnetic field w.r.t.~the line-of-sight. Analogously to the diffusion in a regular magnetic field we can define moments of the diffusion in random IGMFs by
\begin{equation}\label{eq:twodimThetan}
{Y}_{e/\gamma}^{(n)} \equiv \frac{2\pi}{(2^nn!)^2}\int\limits_{0}^\infty{\rm d}\theta\,\theta\,\theta^{2n}\,{\mathcal Y}_{e/\gamma}\quad \text{(random)}\,,
\end{equation}
which follow the same differential equations~(\ref{eq:Thetanevol}) with diffusion matrix ${\mathcal D}_{\rm rnd}$.

So far we have only considered the diffuse scattering of the photons along their initial trajectory. How does this translate into the observed morphology of the $\gamma$-ray signal? Deflections of electrons close to the source at distance $d$, {\it e.g.}~by the magnetic field of the source itself, will have a weaker impact on the observed angular distribution than deflections close to the observer. If the cascade experiences a deflection $\Delta \theta$ at a distance $r$ from the observer we can approximate the corresponding angular displacement $\Delta \theta'$ in the observer's frame via $\Delta \theta'/\Delta\theta \simeq (d-r)/d$. We can account for this scaling in the cascade equation by introducing the corresponding scaling in the convection velocity $R_L^{-1}$ or, equivalently, by a scaling of the diffusion matrix of the form ${\mathcal D}' \simeq ((d-r)/d)^2 {\mathcal D}$. In practice, this requires that we repeat the calculation of transfer matrices after sufficiently small propagation distances, for which we then also account for the variation of (differential) interaction rates $\Gamma$ ($\gamma$) with redshift and adiabatic energy loss. With this simple modification the moments ${Y}^{(n)}$ reflect the angular distribution of $\gamma$-ray halos, as long as scattering in the magnetic field is small and the source is emitting isotropically.

\begin{figure}[t]\centering
\includegraphics[height=2.7in]{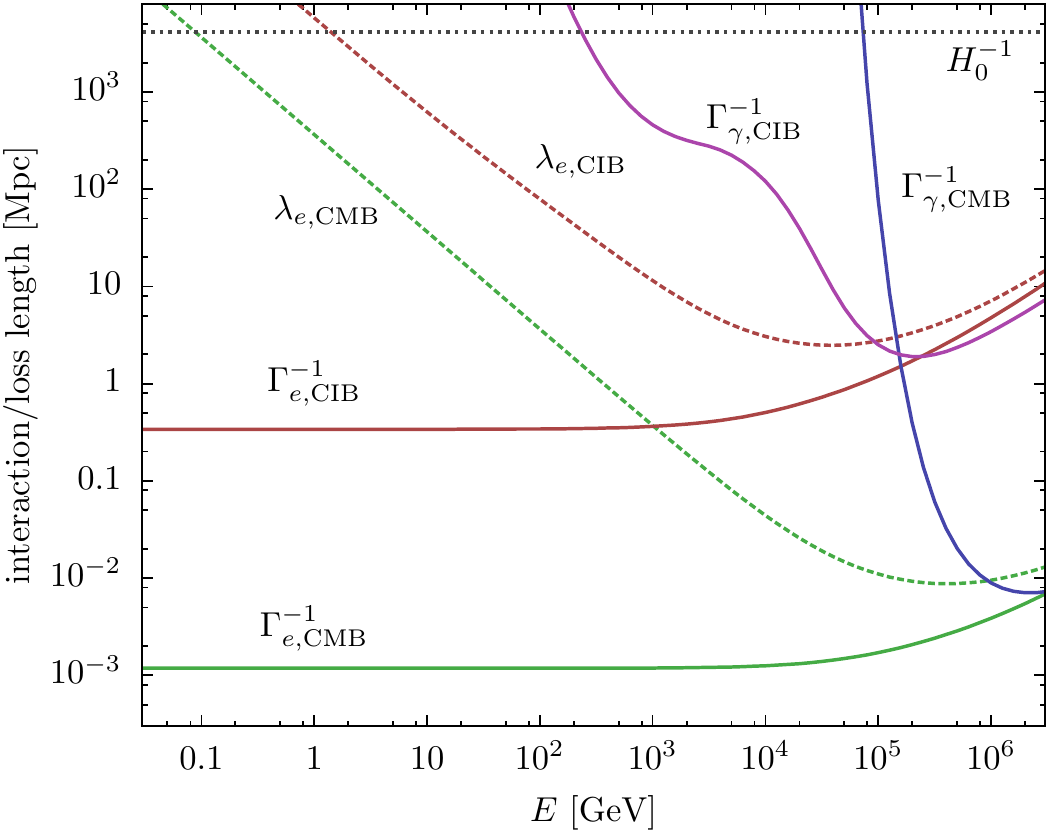}\hfill\includegraphics[height=2.7in]{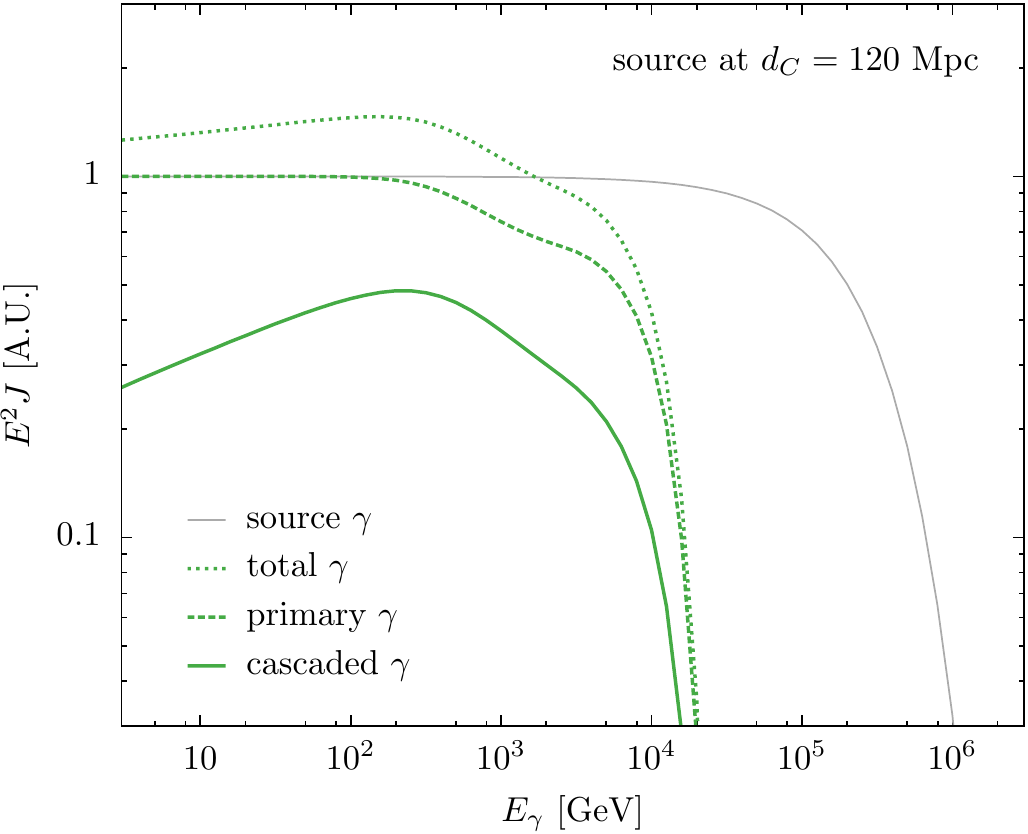}
\caption[]{{\bf Left panel:} The interaction length (solid lines) and energy loss length (dashed lines) from various contributions of the CRB. We show the rates separately for the CMB and CIB. Also shown is the inverse Hubble scale (dotted line). {\bf Right panel:} The spectra of $\gamma$-rays from a source at $120$~Mpc with injection spectrum $Q_\gamma\sim E^{-2}\exp(-E/300\,{\rm TeV})$ (gray line) following Ref.~\cite{Elyiv:2009bx}. We show the contribution of surviving primary $\gamma$-rays (dashed line) and secondary cascaded $\gamma$-rays (solid line) separately.}\label{fig:fig1}
\end{figure}

As an example, we study in the following an isotropic $\gamma$-ray point-source at a distance of about 120~Mpc -- as Mrk 421 -- with a $\gamma$-ray injection spectrum of the form $Q_\gamma \sim E^{-2}\exp(-E/300\,{\rm TeV})$. This particular example has been studied in Ref.~\cite{Elyiv:2009bx} and hence our results are directly comparable. In the left panel of Fig.~\ref{fig:fig1} we show the source spectrum, {\it i.e.}~the spectrum that would be visible without the CRB (thin gray line) together with the electron/positron and $\gamma$-ray spectrum after propagation. The total $\gamma$-ray spectrum (dotted green line) can be decomposed into a ``primary'' component (dashed green line) of surviving $\gamma$-rays and a ``cascaded'' component (solid green line) from $\gamma$-rays of the cascade. The $\gamma$-ray flux is strongly suppressed beyond 10~TeV due to the PP with the CIB and secondary $\gamma$-rays from ICS with the CMB peak between 0.1-1~TeV.

We will assume in the following that the cascade develops in  a weak IGMF with strength $B_0=10^{-15}$~G and a coherence length $\lambda_B=1$~Mpc extends. For the reconstruction of the $\gamma$-ray halo it is convenient to first subtract the moments of the surviving primary $\gamma$-rays that don't take part in the cascade,
\begin{equation}
Y^{(n)}_{\gamma,\,{\rm halo}} = Y^{(n)}_{\gamma,\,{\rm total}} - Y^{(n)}_{\gamma\,{\rm primary}}\,.
\end{equation}
In our example we assume a point-source with sufficiently small angular extend, corresponding to the case $Y^{(0)}_{\gamma,\,{\rm primary}}=Y_{\gamma,\,{\rm primary}}$ and vanishing higher moments. In general, the higher moments of the primary source with an angular extend $2\theta_s$ can be approximated by
\begin{equation}
Y^{(n)}_{\rm primary} \simeq Y^{(0)}_{\rm primary}\frac{\theta_s^{2n}}{n!(n+1)!4^n}\,.
\end{equation}

The size of the first non-trivial moment ${Y}_{\gamma,\,{\rm halo}}^{(1)}/{Y}^{(0)}_{\gamma,\,{\rm halo}}$ already serves as a first indicator for the size of the $\gamma$-ray halo. If this is much larger than the PSF of an IACT the flux of secondary $\gamma$-rays will be strongly isotropized and can only be constrained by the diffuse $\gamma$-ray background (see {\it e.g.}~\cite{Ahlers:2010fw}). We will show in the following that we can use the spectrum of moments to reconstruct the $\gamma$-ray halo for small deflection angles. This will also give an indication at which energies the contribution of secondary $\gamma$-rays contribute to the point-source spectrum.

\section{Reconstruction of the Angular Distribution}\label{sec:reconstruction}

The moments of the $\gamma$-ray halo serve as a measure for its angular distribution. How we can reconstruct the angular distribution from a limited number of moments? As a first step it is convenient to define a distribution $f(E,x)$ by the transformation
\begin{equation}\label{eq:Ndis}
{\mathcal{Y}}_{\gamma, {\rm halo}}(E,\theta)  \equiv Y_{\gamma, {\rm halo}}(E)\int\limits_0^\infty {\rm d}x\left[\frac{1}{(2\pi x)^{\frac{d}{2}}}e^{-\theta^2/2x}\right]f(E,x)\,,
\end{equation}
for regular ($d=1$) or random ($d=2$) magnetic fields. This transformation is motivated by the observation that the kernel $G_d(x,\theta) = [\ldots]$ corresponds to a Green's function of the d-dimensional diffusion  equation, $(\partial_x - \sum_i\partial^2_{\theta_i}) G_d(x,\vec\theta) = 0$ and $G_d(0,\vec\theta)=\prod_i\delta(\theta_i)$ with $\sum_i\theta_i^2 =\theta^2$. We can then identify the quantities ${Y}^{(n)}/{Y}^{(0)}$ as (scaled) moments of the distribution $f(E,x)$ for both, regular and random fields:
\begin{equation}\label{eq:mun}
\mu_n(E) \equiv \int\limits_0^\infty {\rm d}x\,x^n\,f(E,x) = {2^nn!}\frac{{Y}_{\gamma, {\rm halo}}^{(n)}(E)}{{Y}_{\gamma, {\rm halo}}^{(0)}(E)}\,.
\end{equation}

We hence arrive at a classical (Stieltjes) moment problem~\cite{Akhierzer:1965} of finding the distribution $f$ from its moments $\mu_n$. From the differential equations (\ref{eq:Thetanevol}) and the definition (\ref{eq:mun}) it is easy to see that we can find a constant $\mathcal{C}$ such that $\mu_n(E) < {\mathcal C}n![2d\max_{E'\geq E}(\mathcal{D}(E',E))]^n$, where $d$ the distance to the source. This is a sufficient condition for a determinate moment problem~\cite{Akhierzer:1965}, {\it i.e.}~there exists a unique solution $f$ satisfying Eq.~(\ref{eq:mun}). Note, that the reconstruction of $f$ from the {\it complete} set of moments $\mu_n$ is trivial. For instance, we can express $f$ by an infinite sum of Laguerre polynomials which are orthogonal on $[0,\infty)$ under the measure $\exp(-x)$. However, this method does not prove convenient if there are only a finite number of $\mu_n$ at our disposal. The truncation of the expansion after the first $N+1$ basis function leads typically to rapidly oscillating solutions.  Alternatively,  we can reconstruct the distribution by a sequence of approximations $f$, which are maxima of an entropy functional~\cite{Mead:1983qg}, where the condition (\ref{eq:mun}) are introduced via Lagrange multipliers. This problem can then be reduced to a minimization problem of an $N$-dimensional effective potential.

In our case we choose a different approach, which is suitable for the particular form of the distribution. First, we introduce the Laplace transform of the potential $f$ as
\begin{equation}\label{eq:genfunction}
\hat f(E,s) = \mathcal{L}\lbrace f(E,x)\rbrace\equiv \int\limits_0^\infty {\rm d}x e^{-sx}f(E,x) = \sum\limits_{k=0}^\infty\frac{(-s)^k}{k!}\mu_k\,.
\end{equation}
The Laplace transform $\hat f$ corresponds to a generating function of the moments, $(-1)^n\partial^n_sf(E,s)|_{s=0} = \mu_n(E)$, and the solution to the moment problem corresponds to the inverse Laplace transform $f(E,x) = \mathcal{L}^{-1}\lbrace\hat f(E,s)\rbrace$. However, in practice we have only a finite number of moments $N+1$ and the truncation of the alternating series~(\ref{eq:genfunction}) does not converge for large $s$. 

\begin{figure}[t]\centering
\includegraphics[height=2.7in]{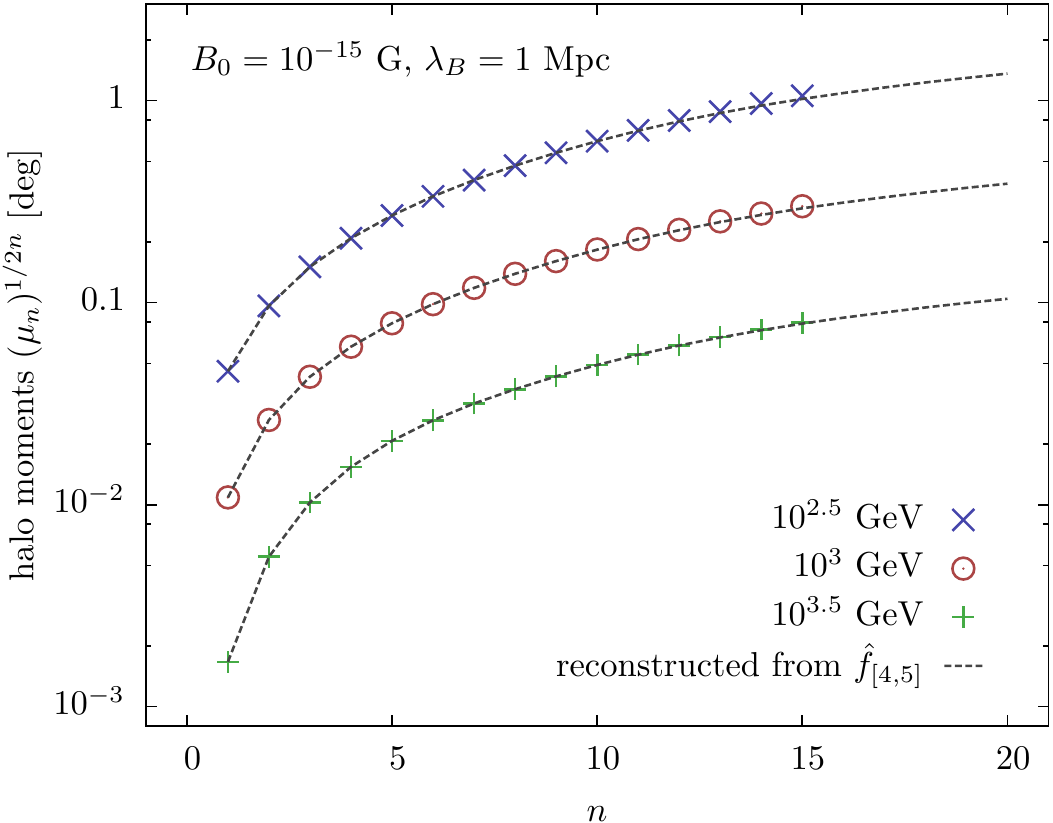}\hfill\includegraphics[height=2.7in]{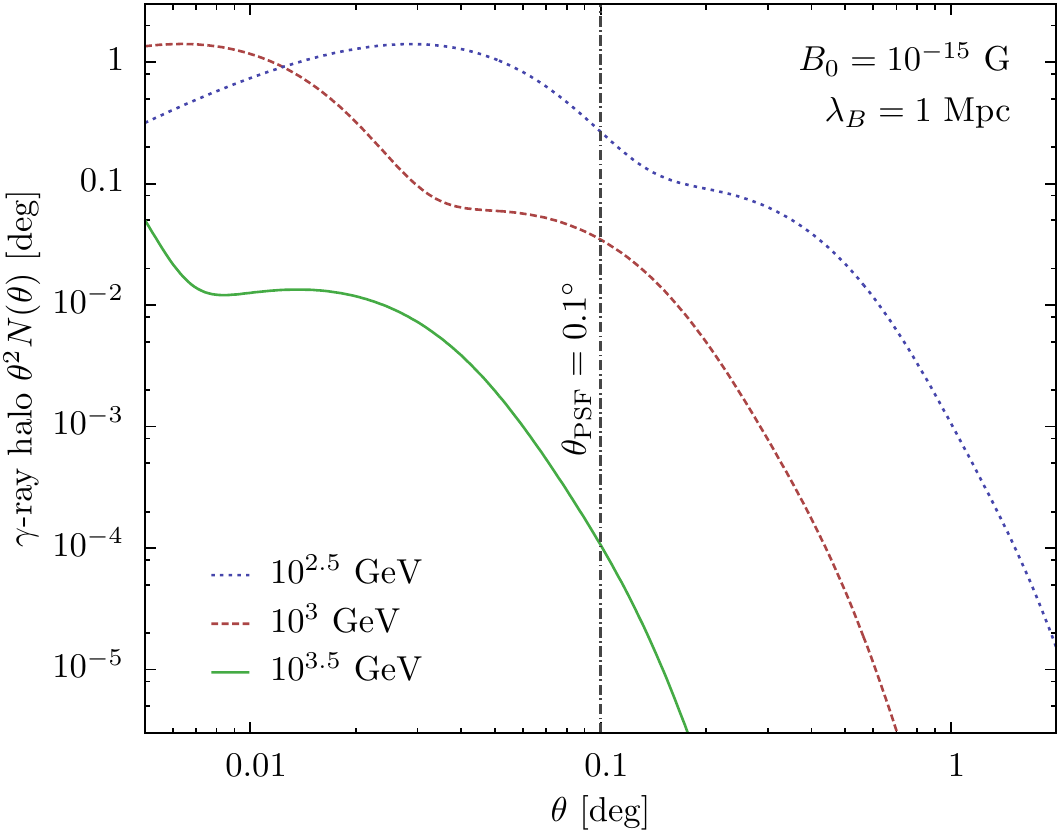}
\caption[]{{\bf Left panel:} The first 15 non-trivial moments $\mu_n$ at three different $\gamma$-ray energies for the example shown in the right panel of Fig.~\ref{fig:fig1} and assuming a IGMF with strength $B_0=10^{-15}$~G and $\lambda_B=1$~Mpc. The dashed lines show the momenta reconstructed by the Pad\'e approximations $\hat f_{[4,5]}$ that are fixed by the first 6 calculated moments. The approximation reproduces the higher moments well. {\bf Right panel:} The $\gamma$-ray halos reconstructed from the moments shown in left panel. We also indicate the typical size of the PSF for IACTs.}\label{fig:fig2}
\end{figure}

We can find an approximate solution by replacing the truncated series by a Pad\'e approximation -- a method which is well-known to chemistry, engineering or nuclear physics~\cite{Baker:1981}. We are approximating $\hat f$ by a rational function $\hat f_{[M,M+1]}(s) = P(s)/Q(s)$ where $P$ and $Q$ are polynomials of degree $M$ and $M+1$, respectively. The coefficents of $P$ and $Q$ are determined by matching the first $2M+1$ terms of the Taylor expansion of $f_{[M,M+1]}$ to the truncated series. Clearly, for $N+1$ calculated moments we can only consider $M\leq N/2$ for the approximation. Since $\deg(Q)>\deg(P)$ the Pad\'e approximation is finite as $s\to \infty$, in contrast to the truncated series it approximates. If we write the denominator via its roots $s_i$ with multiplicity $m_i$, $Q(s) = \prod_{i=1}^n(s-s_i)^{m_i}$, the inverse Laplace transform of the rational function $f_{[M,M+1]}$ has the simple form
\begin{equation}\label{eq:fapprox}
f(E,x) \simeq \mathcal{L}^{-1}\lbrace\hat f_{[M,M+1]}\rbrace = \sum_{i=1}^{n}\sum_{j=1}^{m_i}\frac{c_{ij}(E)}{(j-1)!}x^{j-1}e^{xs_i(E)}\,,
\end{equation}
where the coefficients $c_{ij}$ follow from an expansion into partial fraction. Note, however, that for a general Pad\'e approximation it is not guaranteed that all $\Re(s_i)<0$ and hence the approximation (\ref{eq:fapprox}) can be unstable even if the exact solution (\ref{eq:genfunction}) is stable itself. However, by lowering the degree of approximation $M$ it is in general possible to obtain a stable Pad\'e approximation that fulfills the necessary criteria. This can be done by trial and error -- as we do here for simplicity -- or by an algorithmic procedure~\cite{Hutton}. We will show in the following section, that this procedure is stable and reproduces the moments of the distribution well. Finally, the distribution $N(\theta)$ can be obtained from Eqs.~(\ref{eq:Ndis}) and (\ref{eq:fapprox}).

We illustrate this procedure for the cascade spectrum shown in the right panel Fig.~\ref{fig:fig1}. In the left panel of Fig.~\ref{fig:fig2} we show the first 15 non-trivial moments $\mu_n$ of the distribution $f$ for $\gamma$-ray halos at $10^{2.5}$, $10^3$ and $10^{3.5}$~GeV. The dashed line shows the moments calculated via the Pad\'e approximation $\hat{f}_{[4,5]}$. Note, that this approximation is determined by the first eight non-trivial moments, but also reproduces all the higher moments of our calculation satisfactorily. Using Eqs.~(\ref{eq:Ndis}) and (\ref{eq:fapprox}) we can derive the angular distribution of the halos which are shown in the right panel of Fig.~\ref{fig:fig2}. For illustration we normalize the distribution as $N(\theta,E)={\mathcal Y}_{\gamma,{\rm halo}}(\theta,E)/Y_{\gamma,{\rm total}}(E)$. Note, that not all of this $\gamma$-ray halo will be resolvable in IACTs. We are indicating in the plot the typical size of the PSF of $0.1^\circ$. We will discuss in the following section the size of these $\gamma$-ray halos in more detail.

\begin{figure}[t]\centering
\includegraphics[height=2.7in]{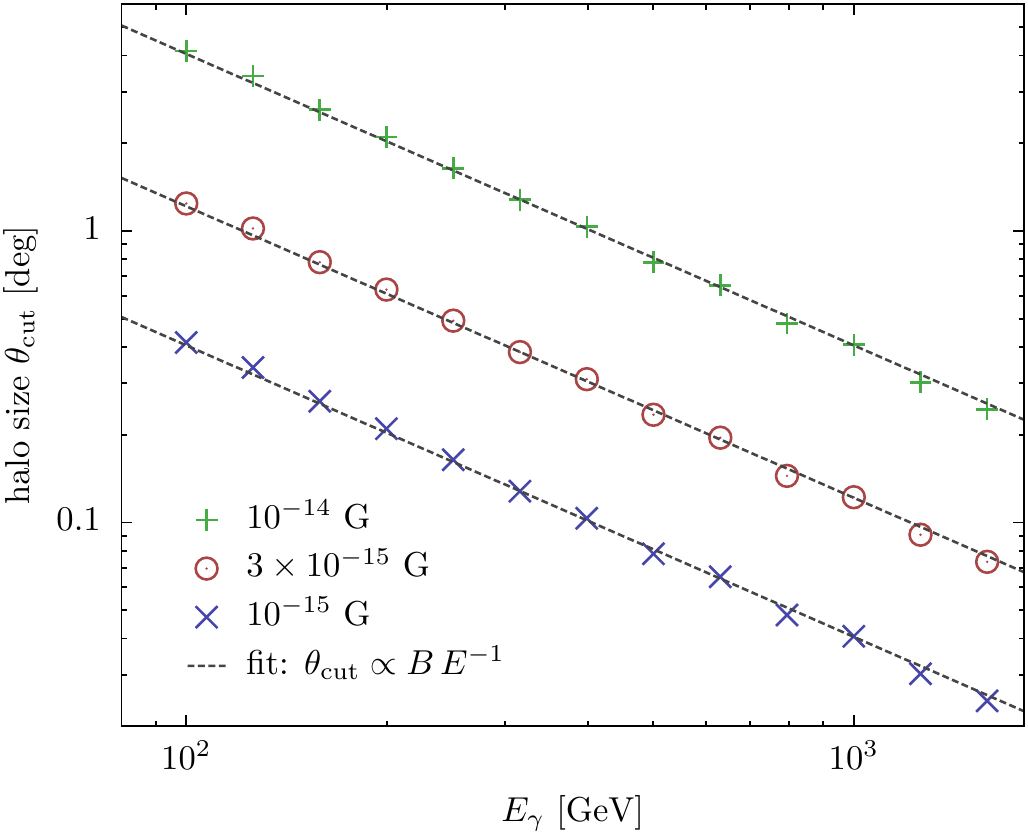}\hfill\includegraphics[height=2.7in]{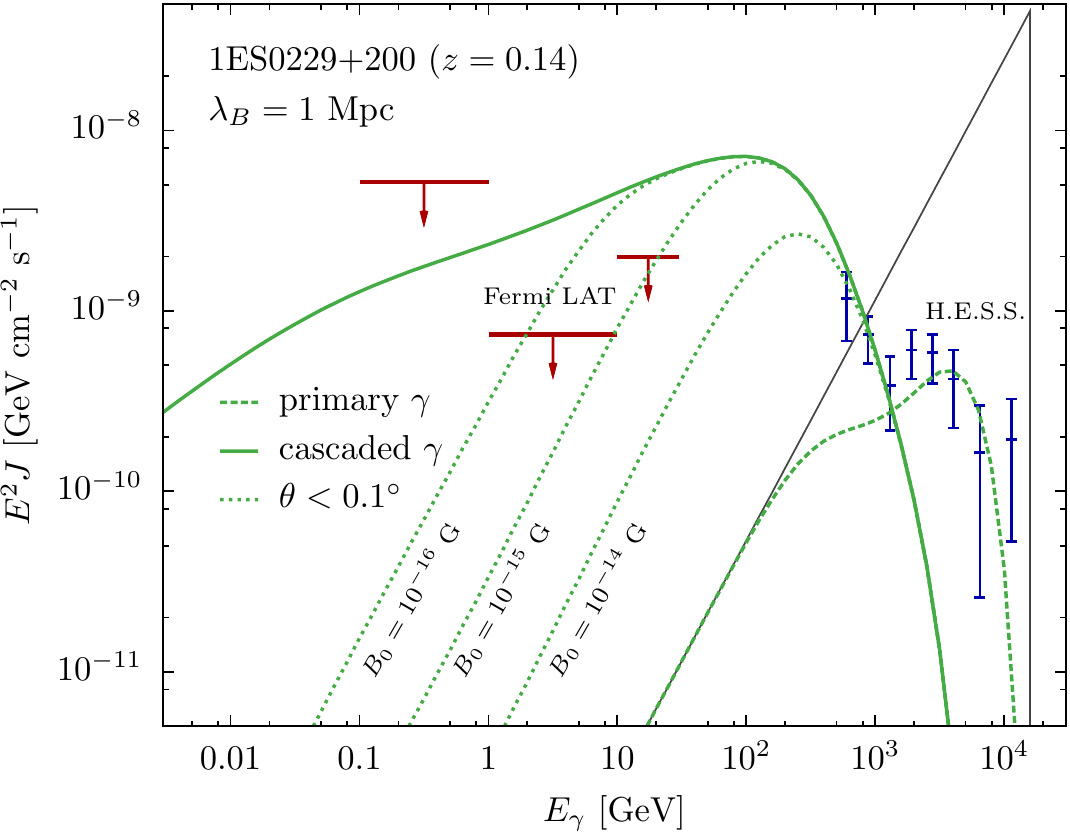}
\caption[]{{\bf Left panel:} The size of the extended $\gamma$-ray halo defined by Eqs.~(\ref{eq:Nasym}) and (\ref{eq:numeric}) for a source at $z=0.031$ with spectrum $Q_\gamma(E) \sim E^{-2}\exp(-E/300{\rm TeV})$. {\bf Right panel:} A model for the $\gamma$-ray spectrum of the blazar source 1ES0229+200 located at $z=0.14$. The blue data points show the H.E.S.S.~observation~\cite{Aharonian:2007wc} and the red lines correspond to the upper flux limits from Fermi-LAT (taken from Ref.~\cite{Tavecchio:2010mk}). We assume a source spectrum of the form $Q_\gamma \propto E^{-2/3}\Theta(20{\rm TeV}-E)$. The solid green line shows the spectrum of secondary $\gamma$-rays without deflections in the IGMF. The dotted green lines indicate the part of the cascaded $\gamma$-ray spectrum within $0.1^\circ$ around the source for an IGMF with coherence length $\lambda_B=1$~Mpc and strength $B_0=10^{-16}$~G, $10^{-15}$~G and $10^{-14}$~G, respectively. }\label{fig:fig3}
\end{figure}

\section{Size of the Extended Halos}\label{sec:example}

The size of the extended halo serves as a measure of the IGMF. Typically, the low-$\theta$ form of the halo derived from the approximation~(\ref{eq:fapprox}) depend on a few roots $s_i$ with large real component $|\Re(s_i)|$. In this case, the $\theta$-distribution is in the form of a modified Bessel function for a random IGMF with $\lambda_B\ll d$. The sub-halos have the form
\begin{equation}\label{eq:Nasym}
N(\theta) \sim \frac{|\Re(s_i)|}{\pi}K_0(\sqrt{2|\Re(s_i)|}\theta) \sim  \frac{1}{\sqrt{8\pi\theta}}\frac{1}{\theta_i^{3/2}}e^{-\theta/\theta_i}\,,
\end{equation}
where in the last step we took the asymptotic form of $K_0$ at large $\theta$ and introduced the characteristic size of the sub-halo, $\theta_{i}=(2|\Re(s_i)|)^{-1/2}$. Hence, there is a simple relation between the measurable size of the halo and the simple poles of the Pad\'e approximation.

In the cases shown in the right panel of Fig.~\ref{fig:fig2} the leading order halo is below the typical instrument's resolution of $\theta_{\rm PSF}=0.1^\circ$. Instead, the next-to-leading-order root will determine the size of the halo. In general, we hence define the size of the leading (observable) halo as
\begin{equation}\label{eq:numeric}
\theta_{\rm cut} ={\rm min}(\lbrace\theta_i\rbrace | \theta_i>\theta_{\rm PSF})\,.
\end{equation}
For our test spectrum we show the parameter $\theta_{\rm cut}$ in the left panel of Fig.~\ref{fig:fig3} for various magnetic field strengths and $\gamma$-ray energies between $100$~GeV to a few TeV. As before we consider a coherence length of $\lambda_B=1$~Mpc. The size of the halo in this energy range follows approximately $\theta_{\rm cut} \propto E^{-1}$ as the fit shows. This agrees with the findings of Ref.~\cite{Elyiv:2009bx} (Fig.~7) derived from a Monte-Carlo study.

Another interesting situation occurs if the cascaded $\gamma$-ray spectrum dominates over the primary $\gamma$-ray emission. This can happen for injection spectra that are considerably harder than $E^{-2}$. In this case the detection sensitivity of the cascaded GeV-TeV spectrum depends on the size of the halo and the resolution of the telescope. As an example we consider the emission of the blazar source 1ES0229+200 located at redshift $z=0.14$, which has been detected by its TeV $\gamma$-ray emission by H.E.S.S.~\cite{Aharonian:2007wc}. The spectrum is shown in the right panel of Fig.~\ref{fig:fig3} as the blue data. Following Ref.~\cite{Tavecchio:2010mk} we model the $\gamma$-ray emission spectrum as $Q_\gamma \propto E^{-2/3}\Theta(20{\rm TeV}-E)$ (thin gray line). The surviving primary $\gamma$-rays are shown as a dashed green line and secondary cascaded $\gamma$-rays by a solid line. 

It is easy to understand the shape of the various spectra. Primary $\gamma$-rays close to $E_{\rm max}$ interact with the CIB to produce electron/positron pairs. This is a slow process happening on typical scales of the order of a few 100~Mpc (see left panel of Fig.~\ref{fig:fig1}). The leptons quickly lose energy via ICS with CMB photons at a rate $b_{\rm ICS} = E/\lambda_e$; their spectrum in quasi-equilibrium ($\partial_tY_e\simeq 0$) follows the differential equation $\partial_E(b_{\rm ICS}Y_e) \simeq \Gamma_{\rm PP}Y_\gamma$. Thus, the Comptonized electron spectrum for $E\ll E_{\rm max}$ has the form $Y_e\sim E_e^{-2}$. The typical photon energy from ICS of a background photon with energy $\epsilon$ is given by $E_\gamma \simeq \epsilon(E_e/m_e)^2$. The resulting photon spectrum at $E\ll E_{\rm max}$ follows from energy conservation in ICS, $\partial_t Y_\gamma \simeq ({\rm d}E_e/{\rm d}E_\gamma)(b_{\rm ICS}/E_\gamma)Y_e\sim (E_e/E_\gamma)^2/(2\lambda_e)Y_e \sim E_\gamma^{-3/2}$. The plateau of the full cascaded spectrum shown in the right panel of Fig.~\ref{fig:fig3} is slightly softer than this since a part of the inverse-Compton spectrum is still above pair-production threshold and enters a second cascade cycle. 

We also show the expected contribution of secondary $\gamma$-rays confined within the PSF of a typical IACT with $\theta_{\rm PSF}=0.1^\circ$ assuming an IGMF with coherence length $\lambda_B=1$~Mpc and strength $B_0=10^{-16}$, $B_0=10^{-15}$ and $B_0=10^{-14}$~G, respectively. The deflection of an electron of the Comptonized spectrum is approximately $\theta_e \sim \lambda_{e}/R_L/4$ following from $\partial_t(\theta_e Y_e) \simeq Y_e/R_L$ and $Y_e\sim E_e^{-2}$. This is consistent with the results of our diffusion ansatz since the first moment of the electron/positron distribution follows $Y^{(1)}_e/Y^{(0)}_e \simeq (\lambda_e/R_L/4)^2/2$. If the typical deflection $\theta_e$ exceeds $\theta_{\rm PSF}$ we expect to see a reduction in the point-source flux by a geometric factor $(\theta_{\rm PSF}/\theta_e)^{2} \propto E_e^{-4} \sim E_\gamma^{-2}$. For ICS in the CMB the transition is expected to occur close to the energy
\begin{equation}
E_{\rm cr} \simeq 0.2\,\sqrt{\frac{B_{\rm fG}}{\theta_{{\rm PSF}, 0.1^\circ}}}\,{\rm TeV}\,.
\end{equation}
This agrees well with the reduced cascade flux (``$\theta<0.1^\circ$''; dotted lines) shown in the plot. 

Before we conclude we would like to emphasize a subtlety concerning the contribution of the CIB in ICS. As can be seen from the summary of interaction/loss lengths in the left panel of Fig.~\ref{fig:fig1}, the contribution of the CIB to the total energy loss of ICS is negligible. The $\gamma$-ray spectrum $Y_\gamma$ is hence almost independent of this contribution, but this is not the case for the higher moments $Y^{(n)}_\gamma$. To see this, let us consider a fully Comptonized electron/positron spectrum $Y_e\sim E_e^{-2}$. Following our previous arguments we have $\partial_t (\theta_\gamma Y_\gamma) \sim ({\rm d}E_e/{\rm d}E_\gamma)(b_{\rm ICS}/E_\gamma)\theta_e Y_e \simeq (E_e/E_\gamma)^2/(8R_L)Y_e$ for the Comptonized electron spectrum. The growth of the deflection is hence proportional to $\sqrt{\epsilon}$ and optical photons are expected to contribute much stronger than CMB photons. However, the fraction of photons that contribute with this large deflection is negligible. Inverse-Compton scattering by the CIB will form a shallow plateau of $\gamma$-rays that are negligible for the calculation of the moments of the central halo from the CMB contribution. We can hence neglect this contribution in the calculation of moments which improves the quality of the halo reconstruction at low $\theta$.

\section{Conclusion}\label{sec:conclusion}

We have discussed a novel technique of calculating extended halos of TeV $\gamma$-ray sources in the presence of intergalactic magnetic fields. The method builds on standard cascade equations that account for all particle interactions with the background radiation and treats the effect of secondary electron/positron deflections in intervening magnetic fields by a diffusion ansatz. The moments of the angular distribution can be calculated efficiently by an extended set of cascade equations. The first moments of the distribution already serve as a good estimator of the halo size. We have shown how the full distribution can be reconstructed from further moments via an inverse Laplace transformation of the moment's generating function using Pad\'e approximations. 

Our method applies to situations where the emission of $\gamma$-rays is isotropic or within a sufficiently large jet opening-angle. The $\gamma$-ray halo is expected to show further structure in the more general case. For instance, $\gamma$-ray emission into narrow jets are expected to produce additional breaks in the halo profile~\cite{Neronov:2007zz} and non-spherical geometries in the case of an off-axes emission~\cite{Neronov:2010bd}. For the illustration of the method we have considered a steady $\gamma$-ray emission. In the case of pulsed or short-lived $\gamma$-ray sources there will be a time-delay between primary and secondary $\gamma$-rays due to the increased path length of the leptons. This can also serve as a measure for the intergalactic magnetic field.

\section*{Acknowledgments}

This work is supported by US National Science Foundation Grant No PHY-0969739  and by the Research Foundation of SUNY at Stony Brook.

\appendix

\section{Derivation of Eqs.~(\ref{eq:diffregular}) and (\ref{eq:diffrandom})}\label{sec:appI}

We assume in the following that the magnetic field is perpendicular to the line-of-sight of the source. In this setup electrons and positrons will be deflected by an angle $\theta$ in a plane normal to the magnetic field. For isotropic emission and small deflections the leptons deflected off the line-of-sight are replenished by leptons initially streaming away from the observer. The net effect is a broadening of the $\theta$-distribution (${\mathcal Y}_\pm$) due to a convection term with opposite sign for electrons ($-$) and positrons ($+$). The transport equations take the form
\begin{align}\nonumber
\dot {\mathcal Y}_{\pm}(E,t,\theta) =&\pm\frac{1}{R_L(E)}\partial_\theta {\mathcal Y}_{\pm}(E,t,\theta) - \Gamma_e(E) {\mathcal Y}_{\pm}(E,t,\theta)\\ &+ \int_E{\rm d}E' \bigg(\frac{1}{2}\gamma_{\gamma e}(E',E) {\mathcal Y}_\gamma(E',t,\theta) + \gamma_{ee}(E',E) {\mathcal Y}_\pm(E',t,\theta)\bigg) + \frac{1}{2}{\mathcal L}^\star_e(E,t,\theta)\,.\label{eq:pmdiffregular}
\end{align}
The evolution equation of the total electron/positron cascade ${\mathcal Y}_e = {\mathcal Y}_+ + {\mathcal Y}_-$ can then be written as
\begin{align}\nonumber
\dot  {\mathcal Y}_e(E,t,\theta) =&\frac{1}{R^2_L(E)}\int\limits_0^t{\rm d}t'e^{-\Gamma_e(E)(t-t')}\partial^2_\theta {\mathcal Y}_e(E,t',\theta) - \Gamma_e(E) {\mathcal Y}_e(E,t,\theta) + \mathcal{L}_e^{\rm eff}(E,t,\theta)\\&+\int\limits_0^t{\rm d}t'e^{-\Gamma_e(E)(t-t')}\int_E{\rm d}E'\gamma_{ee}(E',E)\bigg[\dot  {\mathcal Y}_e(E',t',\theta)+ \Gamma_e(E') {\mathcal Y}_e(E',t',\theta) -\mathcal{L}_e^{\rm eff}(E',t',\theta)\bigg]\,,\label{eq:masterregular}
\end{align}
with an effective source term
\begin{equation}
\mathcal{L}_e^{\rm eff}(E,t,\theta) = \sum_{\alpha=e,\gamma}\int_E{\rm d}E' \gamma_{\alpha e}(E',E) {\mathcal Y}_\alpha(E',t,\theta)  + {\mathcal L}^\star_e(E,t,\theta)\,.
\end{equation}
For $t\Gamma_{e}\gg 1$ we can make the replacement $\Gamma_e\exp(-\Gamma_e(t-t'))\to \delta(t-t')$ and Eq.~(\ref{eq:masterregular}) reduces to 
\begin{equation}\label{eq:master2}
\frac{\partial^2_\theta {\mathcal Y}_e(E,t,\theta)}{R_L^2(E)} \simeq\int_E{\rm d}E'\bigg(\Gamma_e(E')\delta(E-E')-\gamma_{ee}(E',E)\bigg)\bigg[\dot  {\mathcal Y}_e(E',t,\theta)+ \Gamma_e(E') {\mathcal Y}_e(E',t,\theta) -\mathcal{L}_e^{\rm eff}(E',t,\theta)\bigg]\,.
\end{equation}
We can further simplify Eq.~(\ref{eq:master2}) by introducing the mean inelasticity,
\begin{equation}
\langle x\rangle= 1 - \int{\rm d}E'\frac{E'}{E}\frac{\gamma_{ee}(E,E')}{\Gamma_{e}(E)}\,.
\end{equation}
The inelasticity of ICS off CMB photons for electron/positron energies below about $100$~TeV is small and we can hence approximate the differential interaction rate by $\gamma_{ee}(E',E) \simeq \Gamma_e(E')\delta(E-E'(1-\langle x\rangle))$. Using this in Eq.~(\ref{eq:master2}) and taking the limit $\langle x\rangle\ll1$ we arrive at 
\begin{equation}
\frac{\partial^2_\theta {\mathcal Y}_e(E,t,\theta)}{\langle x\rangle(E) R_L^2(E)} \simeq-\partial_E\left(E\,\Gamma_e(E)\bigg[\dot  {\mathcal Y}_e(E,t,\theta)+ \Gamma_e(E) {\mathcal Y}_e(E,t,\theta) -\mathcal{L}_e^{\rm eff}(E,t,\theta)\bigg]\right)\,.
\end{equation}
Integrating this equation gives
\begin{equation}
\frac{1}{E\,\Gamma_e(E)}\int_E^\infty{\rm d}E'\frac{\partial^2_\theta {\mathcal Y}_e(E',t,\theta)}{\langle x\rangle(E')R_L^2(E')} \simeq\dot  {\mathcal Y}_e(E,t,\theta)+ \Gamma_e(E) {\mathcal Y}_e(E,t,\theta) -\mathcal{L}_e^{\rm eff}(E,t,\theta)\,.
\end{equation}
We hence arrive at the diffusion term~(\ref{eq:diffregular}) with diffusion matrix (\ref{eq:Dregular}) for a regular magnetic field.

If the coherence length $\lambda_B$ of the magnetic field is smaller than the distant to the source we can not neglect the spatial dependence of the diffusion velocity $R_L^{-1}$. Generalizing to two angular variables $\vec{\theta} = (\theta_1,\theta_2)$ in the plane orthogonal to the line-of-sight we start with
\begin{align}\nonumber
\dot {\mathcal Y}_{\pm}(E,t,\vec\theta) =&\pm\frac{1}{R_L(E)} \vec{n}_{L}(t)\vec\nabla_{\theta} {\mathcal Y}_{\pm}(E,t,\vec\theta) - \Gamma_e(E) {\mathcal Y}_{\pm}(E,t,\vec\theta)\\ &+ \int_E{\rm d}E' \bigg(\frac{1}{2}\gamma_{\gamma e}(E',E) {\mathcal Y}_\gamma(E',t,\vec\theta) + \gamma_{ee}(E',E) {\mathcal Y}_\pm(E',t,\vec\theta)\bigg) + \frac{1}{2}{\mathcal L}^\star_e(E,t,\vec\theta)\,,\label{eq:pmdiffrandom}
\end{align}
where $\vec{n}_L$ is the direction of the Lorentz force projected into the $\theta_1\theta_2$-plane. From here we arrive at
\begin{align}\nonumber
\dot  {\mathcal Y}_e(E,t,\vec\theta) =&\frac{1}{R^2_L(E)}\int_0^t{\rm d}t'e^{-\Gamma_e(E)(t-t')} \vec{n}_{L}(t)\vec\nabla_{\theta}\left[\vec{n}_{L}(t')\vec\nabla_{\theta}{\mathcal Y}_e(E,t',\vec\theta)\right] - \Gamma_e(E) {\mathcal Y}_e(E,t,\vec\theta) + \mathcal{L}_e^{\rm eff}(E,t,\vec\theta)\\&+\int_0^t{\rm d}t'e^{-\Gamma_e(E)(t-t')}\int_E{\rm d}E'\gamma_{ee}(E',E)\bigg[\dot  {\mathcal Y}_e(E',t',\vec\theta)+ \Gamma_e(E') {\mathcal Y}_e(E',t',\vec\theta) -\mathcal{L}_e^{\rm eff}(E',t',\vec\theta)\bigg]\,.\label{eq:masterrandom}
\end{align}
For the evaluation of the second time integral in Eq.~(\ref{eq:masterrandom}) we can proceed as in the case of a regular magnetic field. However, in the first integral we have to account for the fluctuations of $\vec{n}_L(t')$ over the inverse Compton scattering length. These will average to zero except for $\Delta t\lesssim\lambda_B$ and we hence substitute $\Gamma_e\exp(-\Gamma_e(t-t'))\to \min(1,\lambda_B\Gamma_e)\delta(t-t')$. Averaging over the orientation of the magnetic field can be accounted for by an additional factor $1/3$. Proceeding now along the same steps as in the case of a regular field and replacing the angles $\theta_{1/2}$ by spherical coordinates with radius $\theta$ we arrive at the diffusion term~(\ref{eq:diffrandom}) with diffusion matrix~(\ref{eq:Drandom}).

\section{Diffusion-Cascade Equations}\label{sec:appII}

We start from the Boltzmann equations (\ref{eq:boltzmann}) and (\ref{eq:Thetanevol}) and define discrete values ${Y}^{(n)}_{e,i} \simeq \Delta E_i {Y}^{(n)}_{e}(E_i)$, $Q_{e,i} \simeq \Delta E_iQ_e(E_i)$, etc. The combined effect of transitions and deflections within the cascade during a sufficiently small time-step $\Delta t$ can be described by
\begin{align}\label{eq:ThetaCas0}
\begin{pmatrix}{Y}_\gamma({t}+\Delta{t})\\{Y}_e({t}+\Delta{t})\end{pmatrix}^{(0)}_i
  &\simeq \sum_j\begin{pmatrix}T_{\gamma\gamma}(\Delta
  t)&T_{e\gamma}(\Delta t)\\T_{\gamma e}(\Delta
  t)&T_{ee}(\Delta
  t)\end{pmatrix}_{ji}\begin{pmatrix}{Y}_\gamma({t})\\{Y}_e({t})\end{pmatrix}^{(0)}_j + \Delta t\begin{pmatrix}Q_\gamma\\Q_e\end{pmatrix}_i\,,\\\label{eq:ThetaCasn}
\begin{pmatrix}{Y}_\gamma({t}+\Delta{t})\\{Y}_e({t}+\Delta{t})\end{pmatrix}^{(n)}_i
  &\simeq \sum_j\begin{pmatrix}T_{\gamma\gamma}(\Delta
  t)&T_{e\gamma}(\Delta t)\\T_{\gamma e}(\Delta
  t)&T_{ee}(\Delta
  t)\end{pmatrix}_{ji}\begin{pmatrix}{Y}_\gamma({t})\\{Y}_e({t})\end{pmatrix}^{(n)}_j + \Delta t\begin{pmatrix}0&0\\0&{\mathcal D}\end{pmatrix}_{ji}\begin{pmatrix}{Y}_\gamma({t})\\{Y}_e({t})\end{pmatrix}^{(n-1)}_j\quad(n>0)\,,
  \end{align}
The full cascade solution is then given by
 \begin{equation}
\begin{pmatrix}{Y}_\gamma({t}')\\{Y}_e({t}')\end{pmatrix}^{(n)}_i
  \simeq 
  \sum\limits_{m=0}^n\sum_j\mathcal{A}^{(m)}_{ji}(t'-t)\begin{pmatrix}{Y}_\gamma({t})\\{Y}_e({t})\end{pmatrix}^{(n-m)}_j + \Delta t\sum_j\mathcal{B}^{(n)}_{ji}(t'-t)\begin{pmatrix}Q_\gamma\\Q_e\end{pmatrix}_j\,.
\end{equation}
The $2n$ matrizes $\mathcal{A}^{(m)}$ and $\mathcal{B}^{(m)}$ follow the recursive relation 
\begin{align}\label{eq:rec1}
\mathcal{A}^{(n)}(2^p \Delta t) &= \sum\limits_{i=0}^n\mathcal{A}^{(i)}(2^{p-1}\Delta t)\cdot\mathcal{A}^{(n-i)}(2^{p-1}\Delta t)\,,\\
\mathcal{B}^{(n)}(2^p\Delta t) &= \mathcal{B}^{(n)}(2^{p-1}\Delta t) +  \sum\limits_{i=0}^n\mathcal{A}^{(i)}(2^{p-1}\Delta t)\cdot\mathcal{B}^{(n-i)}(2^{p-1}\Delta t)\,,\label{eq:rec2}
\end{align}
where the non-zero initial conditions are $\mathcal{A}^{(0)}(\Delta t) = \mathcal{T}(\Delta t)$, $\mathcal{A}^{(1)}_{ij} = {\rm diag}(0,\Delta t \mathcal{D}_{ij})$ and $\mathcal{B}^{(0)}(\Delta t) =\mathbf{1}$. The matrices $\mathcal{A}^{(0)}$ and $\mathcal{B}^{(0)}$ are the familiar transfer matrices for electro-magnetic cascades in the presence of a source term. Using the recursion relations (\ref{eq:rec1}) and (\ref{eq:rec2}) we can efficiently calculate the matrices $\mathcal{A}^{(n)}$ and $\mathcal{B}^{(n)}$ via matrix-doubling~\cite{Protheroe:1992dx}.


\begin{thebibliography}{99}
   
\bibitem{Kronberg:1993vk}
  P.~P.~Kronberg,
  Rept.\ Prog.\ Phys.\  {\bf 57}, 325 (1994).

\bibitem{Beck:2008ty}
  R.~Beck,
  AIP Conf.\ Proc.\  {\bf 1085}, 83 (2009)
  [arXiv:0810.2923 [astro-ph]].
  
\bibitem{Kulsrud:2007an}
  R.~M.~Kulsrud and E.~G.~Zweibel,
  Rept.\ Prog.\ Phys.\  {\bf 71}, 0046091 (2008)
  [arXiv:0707.2783 [astro-ph]].
  
\bibitem{Grasso:2000wj}
  D.~Grasso and H.~R.~Rubinstein,
  Phys.\ Rept.\  {\bf 348}, 163 (2001)
  [arXiv:astro-ph/0009061].

\bibitem{Widrow:2002ud}
  L.~M.~Widrow,
  Rev.\ Mod.\ Phys.\  {\bf 74}, 775 (2002)
  [arXiv:astro-ph/0207240].

\bibitem{Barrow:1997mj}
  J.~D.~Barrow, P.~G.~Ferreira and J.~Silk,
  Phys.\ Rev.\ Lett.\  {\bf 78}, 3610 (1997)
  [arXiv:astro-ph/9701063].
  
\bibitem{Dolag:2004kp}
  K.~Dolag, D.~Grasso, V.~Springel and I.~Tkachev,
  JCAP {\bf 0501}, 009 (2005)
  [arXiv:astro-ph/0410419].
  
\bibitem{Aharonian:1993vz}
  F.~A.~Aharonian, P.~S.~Coppi and H.~J.~Volk,
  Astrophys.\ J.\  {\bf 423}, L5 (1994)
  [arXiv:astro-ph/9312045].
  
\bibitem{Plaga:1995}
  R.~Plaga, Nature {\bf 374}, 430-432 (1995).
  
\bibitem{Dolag:2009iv}
  K.~Dolag, M.~Kachelriess, S.~Ostapchenko and R.~Tomas,
  Astrophys.\ J.\  {\bf 703}, 1078 (2009)
  [arXiv:0903.2842 [astro-ph.HE]].
  
  \bibitem{Neronov:2007zz}
  A.~Neronov and D.~V.~Semikoz,
  JETP Lett.\  {\bf 85}, 473 (2007)
  [arXiv:astro-ph/0604607].

\bibitem{Elyiv:2009bx}
  A.~Elyiv, A.~Neronov and D.~V.~Semikoz,
  Phys.\ Rev.\  D {\bf 80}, 023010 (2009)
  [arXiv:0903.3649 [astro-ph.CO]].
  
\bibitem{Eungwanichayapant:2009bi}
  A.~Eungwanichayapant and F.~A.~Aharonian,
  Int.\ J.\ Mod.\ Phys.\  D {\bf 18}, 911 (2009)
  [arXiv:0907.2971 [astro-ph.HE]].
  
\bibitem{Murase:2008pe}
  K.~Murase, K.~Takahashi, S.~Inoue, K.~Ichiki and S.~Nagataki,
  Astrophys.\ J.\  {\bf 686}, L67-L70 (2008)
  arXiv:0806.2829 [astro-ph].
  
\bibitem{d'Avezac:2007sg}
  P.~d'Avezac, G.~Dubus and B.~Giebels,
  Astron.\ Astrophys.\  {\bf 469}, 857 (2007)
  [arXiv:0704.3910 [astro-ph]].
  
\bibitem{Neronov:1900zz}
  A.~Neronov and I.~Vovk,
  Science {\bf 328}, 73 (2010).
  [arXiv:1006.3504 [astro-ph.HE]].
  
\bibitem{Dolag:2010ni}
  K.~Dolag, M.~Kachelriess, S.~Ostapchenko and R.~Tomas,
  Astrophys.\ J.\  {\bf 727}, L4 (2011)
  [arXiv:1009.1782 [astro-ph.HE]].

\bibitem{Tavecchio:2010mk}
  F.~Tavecchio, G.~Ghisellini, L.~Foschini, G.~Bonnoli, G.~Ghirlanda and P.~Coppi,
  Mon.\ Not.\ Roy.\ Astron.\ Soc.\  {\bf 406}, L70 (2010)
  [arXiv:1004.1329 [astro-ph.CO]].

\bibitem{Dermer:2010mm}
  C.~D.~Dermer, M.~Cavadini, S.~Razzaque, J.~D.~Finke and B.~Lott,
  [arXiv:1011.6660 [astro-ph.HE]].

\bibitem{Taylor:2011bn}
  A.~M.~Taylor, I.~Vovk and A.~Neronov,
  [arXiv:1101.0932 [astro-ph.HE]].

\bibitem{Neronov:2011ni}
  A.~Neronov, D.~V.~Semikoz and A.~M.~Taylor,
  [arXiv:1104.2801 [astro-ph.HE]].

\bibitem{Protheroe:1992dx}
  R.~J.~Protheroe and T.~Stanev,
  Mon.\ Not.\ R.\ Astron.\ Soc.\ {\bf 264}, 191 (1993).

\bibitem{Blumenthal:1970nn}
  G.~R.~Blumenthal,
  Phys.\ Rev.\  D {\bf 1}, 1596 (1970).

\bibitem{Blumenthal:1970gc}
  G.~R.~Blumenthal and R.~J.~Gould,
  Rev.\ Mod.\ Phys.\  {\bf 42}, 237 (1970).
  
\bibitem{Lee:1996fp}
  S.~Lee,
  Phys.\ Rev.\  D {\bf 58}, 043004 (1998)
  [arXiv:astro-ph/9604098].
  
\bibitem{Nakamura:2010zzi}
  K.~Nakamura {\it et al.}  [Particle Data Group],
  J.\ Phys.\ G {\bf 37}, 075021 (2010).
  
\bibitem{Franceschini:2008tp}
 A.~Franceschini, G.~Rodighiero and M.~Vaccari,
 Astron.\ Astrophys.\  {\bf 487}, 837 (2008)
 [arXiv:0805.1841 [astro-ph]].

\bibitem{Ahlers:2009rf}
  M.~Ahlers, L.~A.~Anchordoqui and S.~Sarkar,
  Phys.\ Rev.\  D {\bf 79}, 083009 (2009)
  [arXiv:0902.3993 [astro-ph.HE]].
  
\bibitem{AhlersSalvado}
  M.~Ahlers and J.~Salvado, in preparation.
  
\bibitem{Ahlers:2010fw}
  M.~Ahlers, L.~A.~Anchordoqui, M.~C.~Gonzalez-Garcia, F.~Halzen and S.~Sarkar,
  Astropart.\ Phys.\  {\bf 34}, 106 (2010)
  [arXiv:1005.2620 [astro-ph.HE]].

\bibitem{Akhierzer:1965}
  N.~I.~Akhierzer, ``The Classical Moment Problem'', Edinburgh: Oliver \& Boyd (1965).
  
\bibitem{Mead:1983qg}
  L.~R.~Mead and N.~Papanicolaou,
  J.\ of Math.\ Phys.\ {\bf 25}, 2404 (1984).

\bibitem{Baker:1981}
  G.A.~Baker, Jr.~and P.~Graves-Morris,``Pad\'e Approximants.'' Parts 1 and 2, London: Addison-Wesley (1981).

\bibitem{Hutton}
  M.~Hutton and B.~Friedland, IEEE Transactions on Automatic Control {\bf 20}, 329-337 (1975).
    
\bibitem{Aharonian:2007wc}
  F.~Aharonian {\it et al.}  [HESS Collaboration],
  Astron.\ Astrophys.\  {\bf 475},L9-L13 (2007)
  [arXiv:0709.4584 [astro-ph]].
  
\bibitem{Neronov:2010bd}
  A.~Neronov, D.~Semikoz, M.~Kachelriess, S.~Ostapchenko and A.~Elyiv,
  Astrophys.\ J.\  {\bf 719}, L130 (2010)
  [arXiv:1002.4981 [astro-ph.HE]].

\end{thebibliography}
\end{document}